\documentclass[aps,pre,onecolumn,oneside,superscriptaddress,showkeys,floatfix,12pt,linenumbers]{revtex4-1}
\usepackage[dvips]{graphicx}  
\usepackage{amsmath,amsfonts}  
\usepackage{latexsym}
\usepackage{color}
\usepackage{wasysym}
\usepackage{array}

\begin{document}
\setrunninglinenumbers
\modulolinenumbers[5]
\linenumbers

\title{Absence of epidemic thresholds in a growing adaptive network}
\author{G\"{u}ven Demirel}
\email[Author for correspondence ]{guven@pks.mpg.de}
\affiliation{Max-Planck-Institute for the Physics of Complex Systems -- N\"{o}thnitzer Stra\ss e 38, 01187 Dresden, Germany}
\author{Thilo Gross}
\affiliation{Merchant Venturer's School of Engineering, University of Bristol -- Woodland Road, Clifton BS8 1UB Bristol, United Kingdom}
\date{\today}

\begin{abstract}
The structure of social contact networks strongly influences the dynamics of epidemic diseases. In particular the scale-free structure of real-world social networks allows unlikely diseases with low infection rates to spread 
    and become endemic. However, in particular for potentially fatal diseases, also the impact of the disease on the social structure cannot be neglected, leading to a complex interplay. Here, we consider the growth of a network by preferential attachment from which 
    nodes are simultaneously removed due to an SIR epidemic. We show that increased infectiousness increases the prevalence of the disease and simultaneously causes a transition from scale-free to exponential topology. Although a transition to a degree distribution 
    with finite variance takes place, the network still exhibits no epidemic threshold in the thermodynamic limit. We illustrate these results using agent-based simulations and analytically tractable approximation schemes.
\end{abstract}


\maketitle

\section{INTRODUCTION}
Throughout human history epidemic diseases have been a constant threat.
The plague of Athens wiped out 25-35 percent of the city's population as early as 430 BC \cite{Hays2005}. A bubonic plague epidemic killed between 75000 and 100000 inhabitants of London between April 1665 and January 1666, as its 
population was about 460000 \cite{Hays2005}. Smallpox became a constant threat to Europe throughout the eighteenth century, where growing cities like London were especially vulnerable to the disease on account of providing incessant susceptible 
immigrant hosts to the virus \cite{Hays2005}. After a brief respite during the 20th century epidemics are on the rise again. For instance in 2010, around 655000 people died of malaria \cite{WHO2011}. Combating such epidemic diseases efficiently in the mega-cities of the 
future is likely to require a heightened understanding of the dynamics of the diseases and the social networks on which they spread.  

The horrendous examples above illustrate that epidemic diseases can have devastating effects on populations. It is easily conceivable that this has a drastic effect on population structure and specifically on the structure of social contact networks in the population. 
Simultaneously, the fate of epidemics strongly depends on properties of the contact networks, leading to a complex interplay. 

In the past two decades, epidemic spreading has been extensively studied on different complex networks to understand the influence of the social contact network structure on the disease prevalence \cite{Keeling2005,Bansal2007,Funk2010}. 
In particular, it has been shown that a crucial determinant of epidemic spreading is the \emph{degree distribution} of the contact network, i.e. the probability distribution of finding an agent with a given number of contacts. If the variance of this distribution is finite then there is generally a threshold of infectiousness, which the disease has to exceed to persist in the network \cite{Moore2000,Pastor-Satorras2001a,Moreno2002}. By contrast, if the variance of the degree distribution is infinite, such as in certain scale-free networks, then the epidemic threshold vanishes
\cite{Pastor-Satorras2001a,Moreno2002,Pastor-Satorras2001b,May2001,Castellano2010} and diseases can persist by hierarchical spreading from hub nodes of high degree \cite{Barthelemy2004}. Thus scale-free networks allow unlikely diseases with low infectiousness to spread and become endemic. 

Today, the structure of social networks is recognized as a key factor with direct implications for epidemic dynamics and potential counter measures  \cite{Pastor-Satorras2002b,Zanette2002,Cohen2003,Holme2004,Meyers2005}.
This insight has motivated the integration of real world network data into epidemic models \cite{Newman2002,Hufnagel2004,Colizza2006,Read2008,Belik2011}. Furthermore, theoretical models 
have been extended by including several properties of real world networks such as degree constraints \cite{Pastor-Satorras2002}, clustering \cite{Eguiluz2002,Read2003,Serrano2006}, information filtering \cite{Mossa2002}, social hierarchy \cite{Grabowski2004},
and nonuniform transmission probabilities \cite{Newman2002b}.  

A relatively recent addition is to consider also the feedback of the epidemic on the social network structure, which can be indirect, e.g. by triggering behavioural changes of agents \cite{Funk2010,Gross2006}, or direct by removing agents due to hospitalization, 
quarantine or death. 

Modelling the network response to an ongoing epidemic leads to an adaptive network, a system in which a dynamical interplay between the dynamics of the network and the dynamics on the network takes place \cite{Gross2008,Gross2009}. Epidemics on adaptive networks 
can exhibit complex emergent dynamics \cite{Gross2006,Shaw2008,Risau-Gusman2009,Graeser2011,Wang2011} (e.g. sustained oscillations, bistability, and hysteresis) and emergent topological properties \cite{Gross2006,Shaw2008,Marceau2010} (e.g. heterogeneous degree 
distributions and assortative degree correlations). Furthermore, the study of the social response to epidemics is interesting from an applied point of view because it could enable enhanced vaccine control \cite{Shaw2010} and effective quarantine strategies 
\cite{Lagorio2011}. 

In comparison to social responses to the epidemics \cite{Gross2006,Shaw2008,Risau-Gusman2009,Graeser2011,Wang2011,Marceau2010}, direct topological feedback via the removal of nodes has received less attention. Previous works considered the case where network growth 
and death processes are balanced and thus the population stays in equilibrium and fluctuates around a fixed system size \cite{Kamenev2008,Schwartz2009}. The case of continuous growth was studied in \cite{Guerra2010}, where new nodes attach preferentially to high degree non-infected nodes. It was observed that a transition from a scale-free topology to an exponential one takes place as the infectiousness decreases. In  another study, network growth and node removals have been incorporated in a single model \cite{Hayashi2004}. However, the authors focused on the epidemic oscillations and did not consider topological effects in detail. Another related work focused on the interplay between network growth and dynamical behaviour in the context of evolutionary game theory \cite{Poncela2009}, where 
new players preferentially attach to those receiving higher payoffs. 

Here, we consider the growth of a network by preferential attachment from which nodes are simultaneously removed due to a susceptible-infected-removed (SIR) epidemic. The appeal of this model lies in its paradoxical nature, in absence of the disease 
preferential attachment will lead to the formation of scale-free topologies in which the epidemic threshold vanishes, such that the disease 
can invade. However, an established SIR disease will quickly infect and remove nodes of high degree such that the variance of the degree distribution is decreased and epidemic thresholds reappear, potentially leading to the extinction of the disease. 

One can attempt to guess the outcome by the following (slightly naive) line of reasoning: Observing a scale-free topology implies 
that the epidemic is extinct. But an extinct epidemic implies scale-free structure and hence vanishing epidemic threshold precluding extinction. Logically, the only possible solution is that the epidemic persists (unconditionally) in a network that is not scale free. In other words, one would expect that the coevolution of epidemic state and network structure should lead to a vanishing epidemic threshold in an exponential network.  

One could argue that the naive reasoning is wrong because the paradox above can be resolved temporally, such that the 
epidemic goes extinct while the network is exponential, whereas a disease-free scale-free network develops at later times. Subsequently the network cannot be re-invaded by the disease because all infected have been removed. However, this temporal resolution is only feasible in finite networks. 
In the thermodynamic limit it can be easily shown that a finite number of infected survive even below the epidemic threshold, which precludes complete extinction and thus is sufficient to reignite the epidemic at a later stage.

In the remainder of this paper we present a detailed dynamical analysis of the epidemic model showing that the naive reasoning 
presented above is actually correct in the thermodynamic limit. Here, high infectiousness elevates the prevalence of the disease but at the same time limits epidemic spreading by causing a transition from scale-free to exponential topologies. Thus a balance is reached where the finite variance of the degree distribution is matched to the finite infectiousness, such that the epidemic persists.
We use a novel analytical approximation scheme to understand this balance in detail, and also identify a small parameter region, where, after all, a temporal resolution of the paradox is observed. 

\section{MODEL}\label{sec:models}
We study the spreading of a susceptible-infected-removed (SIR) disease \cite{Anderson1991} on an evolving network. In this network a given node is either susceptible (state S) or already infected with the disease (state I). We start with a fully connected network 
of $m_{0}$ nodes and consider three dynamical processes: a) the arrival of nodes, b) disease transmission, and c) the removal of nodes. 

In the following we measure all rates per capita, including the arrival rate. This implies that larger populations have a proportionally larger influx of agents. This assumption is necessary to keep the model well-defined in the thermodynamic limit and appears plausible e.g.~for growing cities, where the attractivity increases with size.  

New nodes arrive in the population at a constant rate $q$ and are already infected with the disease with probability $w$. Arriving nodes immediately establish links with $m$ of the nodes selected according to the preferential attachment rule \cite{Barabasi1999}: A new node establishes a link with an existing node of degree k with probability 
$kP_{k}/\langle k \rangle$, where $P_k$ denotes the degree distribution (the probability that a randomly picked node has degree $k$) and $\langle k \rangle = \sum{kP_k}$ is the mean degree. 

Disease transmission occurs at rate $p$ on every link connecting a susceptible and an infected node. Therefore, nodes with higher degree are proportionally more likely to catch and spread the disease.

Removal of infected nodes take place at rate $r$. Because we describe a fatal disease from which recovery is not possible, removed nodes and their links are entirely deleted from the simulation and do not re-enter at a later stage. 

Unless mentioned otherwise, the following set of parameter values is used throughout the paper: $m_0 = 6$, $m = 5$, $q = 0.01$. In agent-based simulations the network is simulated until $N$ reaches $10^7$ (the scale of present mega cities), 
or the network has stopped growing and the time reaches $10^4$.  

\section{ANALYTICAL TREATMENT}\label{sec:approximation}
The dynamics on and of complex networks can be captured by a set of coupled ordinary differential equations, in so-called moment-closure approximations \cite{Marceau2010,Keeling1999,Bauch2005,House2011,DemirelInprep,Noel2009,Lindquist2011,Gleeson2011}. Here, we develop a heterogeneous 
node approximation \cite{Pastor-Satorras2001a,Pastor-Satorras2001b,Moreno2002}, where the network evolution is captured in a set of equations for the node densities in different degree-classes. 
We define $[A_k]$ as the density of nodes of state $A \in {S,I}$ and degree k among all nodes in the network. Furthermore, the total density of S-nodes is denoted as $[S]$, $[S]=\sum_{k}[S_{k}]$. The density $[I]$ is defined analogously such that $[S]+[I]=1$. We note that the number of nodes $N$ is not conserved. Therefore, node densities are renormalized due to node 
arrival and removal events. 

We now formulate a dynamical system that captures the time evolution of the densities $[S_{k}]$ and $[I_{k}]$ due to infection, node arrival, and node removal processes, 
\begin{eqnarray} \label{eq:Sk}
\displaystyle \frac{\rm d}{\rm dt} [S_{k}] & = & \displaystyle q\Bigg((1-w)\delta_{k,m}+\frac{m}{\langle k \rangle}\Big(-k[S_{k}] +(k-1)[S_{k-1}]\Big)-[S_{k}]\Bigg)-p z_{I} [I] k[S_{k}]   \nonumber \\
	   				    &  & \displaystyle + r\Bigg(z_{I} [I]\Big((k+1)[S_{k+1}]-k[S_{k}]\Big) + [I][S_{k}]\Bigg), \nonumber \\
\displaystyle \frac{\rm d}{\rm dt} [I_{k}] & = & \displaystyle q\Bigg(w\delta_{k,m}+\frac{m}{\langle k \rangle}\Big(-k[I_{k}]+(k-1)[I_{k-1}]\Big)-[I_{k}]\Bigg) +p z_{I} [I]k[S_{k}] \nonumber \\
					   &  & \displaystyle + r\Bigg(z_{I} [I]\Big((k+1)[I_{k+1}]-k[I_{k}]\Big)+ [I][I_{k}]-[I_{k}]\Bigg), \ \ \ 0 < k < k_{max} 
\end{eqnarray}

For understanding the equation governing the evolution of the density $[S_k]$, consider that new nodes with degree $m$ arrive at the rate $q$ and have state S with probability $1-w$. Thus, the density $[S_{m}]$ increases at the rate $q(1-w)$. 
A newly arriving node builds a link to a node in the $S_{k}$ class with the probability $k[S_{k}]/\langle k \rangle$ and causes it to pass into the $S_{k+1}$ class. Because $m$ such links are established by each newly arriving node, the density $[S_{k}]$ 
decreases by $qmk[S_{k}]/\langle k \rangle$. Similarly, nodes in the $S_{k-1}$ class pass into the $S_{k}$ class at the rate $qm(k-1)[S_{k-1}]/\langle k \rangle$. Additionally, we need to renormalize the density $[S_{k}]$ when a node arrives. 
This corresponds to a loss of the $[S_{k}]$ density of $q[S_{k}]$.

At rate $p$, nodes within the $S_{k}$ class become infected through their links with infected nodes causing them to pass into the $I_{k}$ class. The density of such links can be approximated to good precision by $z_{I}[I]k[S_{k}]$, where $z_{I}=\langle k_{I} \rangle / \langle k \rangle$.

Finally, nodes within the $S_{k}$ class pass into the $S_{k-1}$ class due to the removal of their infected neighbours. Given the density of such links, $z_{I}[I]k[S_{k}]$, the density of $[S_{k}]$ decreases by $rz_{I}[I]k[S_{k}]$. Similarly, nodes in the 
$S_{k+1}$ class pass into the $S_{k}$ class corresponding to a gain of \ $rz_{I}[I](k+1)[S_{k+1}]$. As infected nodes are removed, the density of all degree classes increases due to the renormalization leading to a gain of $r[I][S_{k}]$ for the density $[S_{k}]$.
The rate equation for the density $[I_ {k}]$ is constructed analogously.

In the following we refer to equation~(\ref{eq:Sk}) as the heterogeneous approximation. The main drawback of such heterogeneous approximations is the high dimensionality of the system of equations, which complicates the analytical solution and thus typically 
necessitates extensive numerical studies. Furthermore, since it is not possible to numerically integrate an infinite dimensional system of differential equations, we need to introduce a degree cut-off $k_{max}$ by assuming 
$ \sum_{k=k_{max}+1}^{\infty}{P_{k}} \ll \sum_{k=0}^{k_{max}}{P_{k}}$. The higher the degree cut-off, $k_{max}$, the more precise the heterogeneous approximation becomes. 

In the following, we develop a low dimensional approximation by summing over the degree classes. We consider the susceptible proportion of the population $[S]$, the mean degree $\langle k \rangle$, and the mean degree of susceptibles $\langle k_S \rangle$ which evolve according to
\begin{eqnarray}\label{eq:cons}
\displaystyle \frac{\rm d}{\rm dt} [S] & = & \displaystyle \sum_{k} \frac{\rm d}{\rm dt} [S_{k}], \nonumber \\
\displaystyle \frac{\rm d}{\rm dt} \langle k \rangle & = & \displaystyle \sum_{k} k\frac{\rm d}{\rm dt} ([S_{k}]+[I_{k}]),\nonumber \\
\displaystyle \frac{\rm d}{\rm dt} \langle k_{S} \rangle [S] & = & \displaystyle \sum_{k} \frac{\rm d}{\rm dt} k[S_{k}].
\end{eqnarray}

Using equation~(\ref{eq:Sk}) and equation~(\ref{eq:cons}), we obtain
\begin{eqnarray}\label{eq:low_hetero}
\displaystyle \frac{\rm d}{\rm dt} [S] & = &  \displaystyle q(1-w-[S])-p \frac{\langle k_{S} \rangle \langle k_{I} \rangle}{\langle k \rangle}[S][I] +r[S][I], \nonumber \\
\displaystyle \frac{\rm d}{\rm dt} \langle k \rangle & = &  \displaystyle q(2m-\langle k \rangle) + r(2\langle k_{S} \rangle[S] -\langle k \rangle(1+[S])), \nonumber \\
\displaystyle \frac{\rm d}{\rm dt} \langle k_{S} \rangle & = & \displaystyle q\left( \frac{(1-w)(m - \langle k_{S} \rangle)}{[S]} + m \frac{ \langle k_{S} \rangle}{\langle k \rangle}\right)- p \frac{\langle k_{I} \rangle [I]}{\langle k \rangle} (\langle k_{S}^2 \rangle-\langle k_{S} \rangle ^2) -r\frac{\langle k_{S} \rangle\langle k_{I} \rangle}{\langle k \rangle}[I].
\end{eqnarray}
 
In the following, we refer to equation~(\ref{eq:low_hetero}) as the coarse-grained heterogeneous approximation. 

Because we have not derived an equation for the second moment of the 
susceptible degree distribution $\langle k_{S}^2 \rangle$, equation~(\ref{eq:low_hetero}) does not constitute a closed dynamical system. We address this problem by replacing $\langle k_{S}^2 \rangle$ by $\langle k_{S} \rangle^2+\langle k_{S} \rangle$ in an additional 
approximation. We note that this approximation is valid exactly when the network has a Poissonian degree distribution. It can therefore be thought of as a \textquoteleft random-graph approximation\textquoteright. This approximation will certainly fail in 
the case of scale-free degree distribution because of the degree distributions with diverging variance, i.e. $\langle k_{S}^2 \rangle\rightarrow\infty$. 
By contrast, as will become apparent below, the approximation still performs well for distributions with large finite variance. 

Using the random-graph approximation we obtain 
\begin{eqnarray}\label{eq:S}
\displaystyle \frac{\rm d}{\rm dt} [S] & = & \displaystyle q(1-w-[S])-p \frac{\langle k_{S} \rangle \langle k_{I} \rangle}{\langle k \rangle}[S][I]+ r[S][I], \nonumber \\
\displaystyle \frac{\rm d}{\rm dt} \langle k \rangle & = & \displaystyle q(2m-\langle k \rangle) + r(2\langle k_{S} \rangle[S]-\langle k \rangle(1+[S])), \nonumber \\
\displaystyle \frac{\rm d}{\rm dt} \langle k_{S} \rangle & = & \displaystyle q\left( \frac{(1-w)(m - \langle k_{S} \rangle)}{[S]} + m \frac{ \langle k_{S} \rangle}{\langle k \rangle}\right)- p \frac{\langle k_{S} \rangle \langle k_{I} \rangle [I]}{\langle k \rangle}-r\frac{\langle k_{S} \rangle\langle k_{I} \rangle}{\langle k \rangle}[I],
\end{eqnarray}
where $[I]$ and $\langle k_{I} \rangle$ are given by the conservation laws, such that the system constitutes a closed model. In the following we refer to this model as the homogeneous approximation.

\section{BASIC ANALYSIS}\label{sec:analysis}
In this section, we present a detailed analysis of the analytical model and confirm the results by comparison with agent-based simulations of the network.

Before we launch into a detailed discussion of the model, let us consider the limiting case of network evolution in the absence of the epidemic. In this case the model is identical to the Barab\'{a}si-Albert model of network growth \cite{Barabasi1999}, which is known 
to lead to scale-free topologies, where the degree distribution follows a power law $P_{k}\propto k^{-\gamma}$ with exponent $\gamma=3$. In the present model the emergence of scale-free topologies is thus expected in the limit where the disease goes extinct or remains limited to a finite 
number of infected nodes $\ll N$. When the epidemic is present, high degree nodes are disproportionately likely to become infected and subsequently removed, which can be expected to prevent the formation of scale-free topologies. 

\begin{figure}[ht!]
  \centering
   \includegraphics[width=100mm,keepaspectratio]{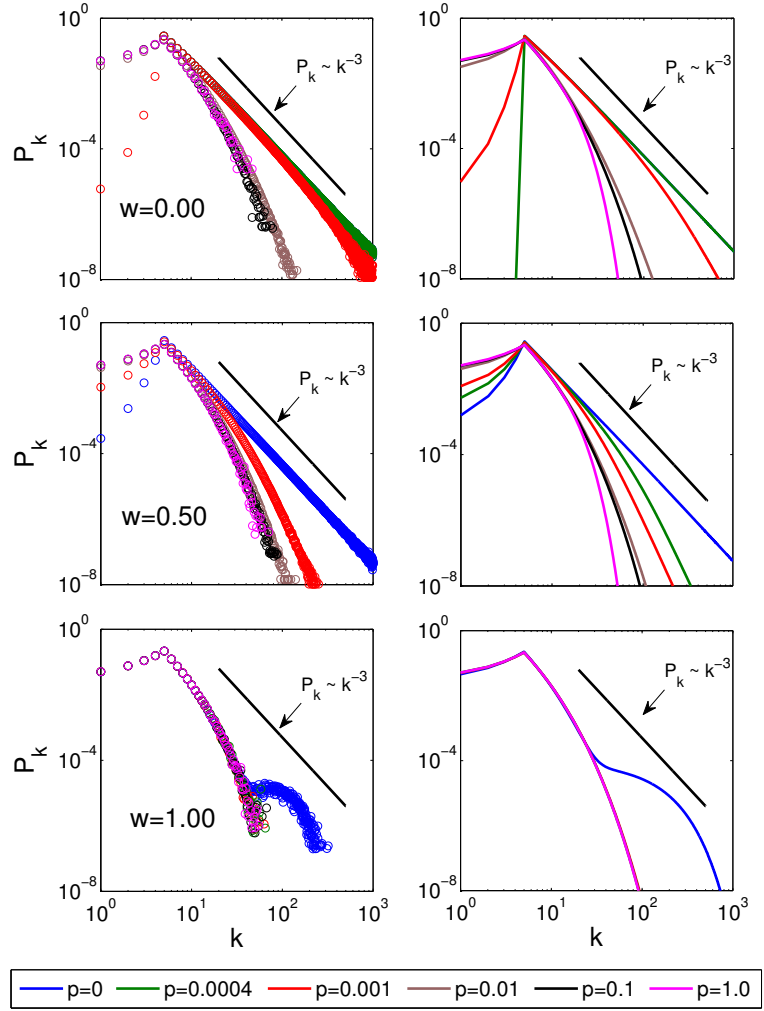}
  \caption{Degree distribution for varying infectiousness $p$ and infected arrival fraction $w$. Left: individual-based simulations. Right: heterogeneous approximation. Scale-free degree distributions are observed when disease 
infectiousness vanishes ($p=0$). For increasing infectiousness the degree distribution quickly becomes exponential. Parameters: $r = q = 0.01$, $m_0 = 6$, $m= 5$. Shown are averages over 1000 simulation runs.}
  \label{fig:degreedist}
\end{figure}

We confirm this intuition by plotting degree distributions for various parameter sets in figure~\ref{fig:degreedist}. Because the homogeneous approximation does not provide any information on the degree distribution, we show a comparison of the heterogeneous approximation 
with agent-based simulations. The figure shows a good agreement between the modelling approaches and confirms basic intuition. When all arriving nodes are susceptible ($w =0$), a scale-free degree distribution with the expected exponent $\gamma=3$ is formed for $p=0$. 
At finite infectiousness $p$, the topology changes from scale-free to exponential. The same behaviour is observed at higher rates of infected arrivals, $0<w<1$.

When all arriving nodes are already infected ($w=1$), the distribution has a bimodal form for $p=0$ with high degree contribution coming from the initial susceptibles which never get infected. At positive infectiousness $p$, these individuals 
eventually die and the mode at high degrees disappears. 

\begin{figure}[ht!]
  \centering
  \includegraphics[width=100mm,keepaspectratio]{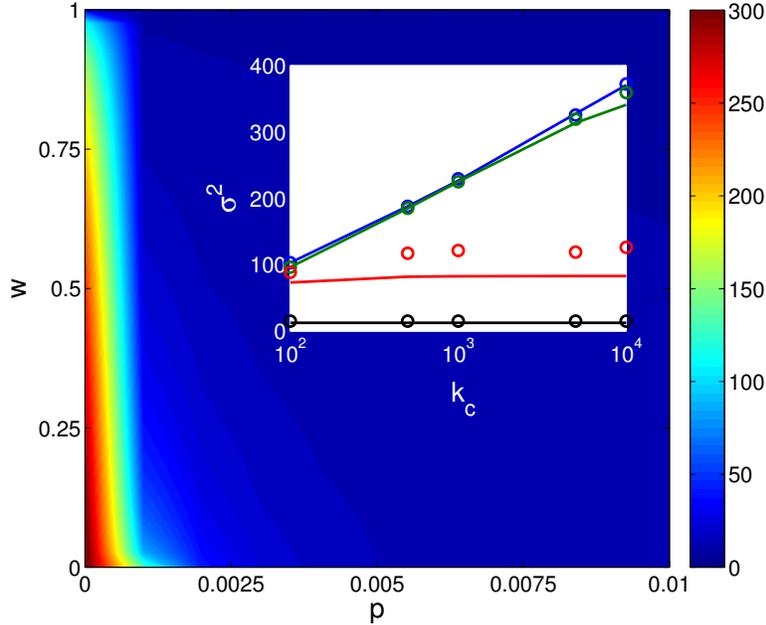}
  \caption{Variance $\sigma^{2}$ of the degree distribution. Degree variance $\sigma^{2}$ is plotted as a function of the infectiousness $p$ and the infected arrival fraction $w$ from agent-based simulations. The degree variance $\sigma^{2}$ decreases with increasing infectiousness $p$ and fraction of infected arrivals $w$. Inset: Degree variance $\sigma^{2}$ as a function of the degree cut-off $k_{c}$. A transition from a cut-off dependent ($p=0$, scale-free) to an independent ($p=0.0004$, $p=0.001$, and $p=0.005$, exponential) regime is observed as the infectiousness $p$ is increased. In network simulations (circles), nodes were restricted to at most $k_{c}$ neighbours. 
In the heterogeneous approximation (solid lines) the cut-off $k_{c}$ is directly imposed as $k_{\rm max}$. Parameters: $r = q = 0.01$, $m_0 = 6$, $m= 5$, $k_c = 5\times10^{3}$.}
  \label{fig:stddev}
\end{figure}

In order to quantify the topological transition from the scale-free to the exponential degree distribution, we plot the variance $\sigma^{2}$ of the degree distribution as a function of infectiousness $p$ and fraction of infected arrivals $w$ in figure~\ref{fig:stddev}. As either parameter increases, the disease prevalence in the steady-state, $[I]^*$, increases and removal occurs at a high rate. As a result the degree distribution becomes narrower and the degree variance decreases. It is apparent that very high values of the variance are 
only found for low infectiousness $p$, whereas higher infectiousness quickly lead to narrow distributions. 

Above we computed the variance $\sigma^{2}$ of the degree distribution of networks. One concern in any computation of this kind is finite-size effects. In the heterogeneous approximation these effects appear directly in form of the maximal degree that is considered in the approximation. In agent-based simulation a similar cut-off exists as the maximal degree in a network of finite size is bounded by the number of nodes. Hence all moments of the degree distribution, including the variance, must be finite regardless of the shape of the degree distribution. However, if the finite networks are drawn from an ensemble that becomes scale-free in the thermodynamic limit
the variance $\sigma^{2}$ is often found to increase logarithmically with the imposed cut-off \cite{Pastor-Satorras2002}. 

We now rule out that low values of the degree variance $\sigma^{2}$, observed above, were due to finite-size effects by considering the variance $\sigma^{2}$ as a function of the degree cut-off $k_{c}$ (figure~\ref{fig:stddev} inset). For the case of $p=0$, 
where we observed scale-free behaviour, we find that the observed variance increases logarithmically as expected. Conversely, for the finite values of infectiousness $p$ the observed $\sigma^{2}$ is insensitive to a sufficiently large cut-off. In summary these results show that 
fatal diseases should relatively quickly destroy scale-free structure of social networks at all but the smallest removal rate and/or infectiousness.

\begin{figure}[ht!]
  \bigskip 
  \centering
  \includegraphics[width=100mm,keepaspectratio]{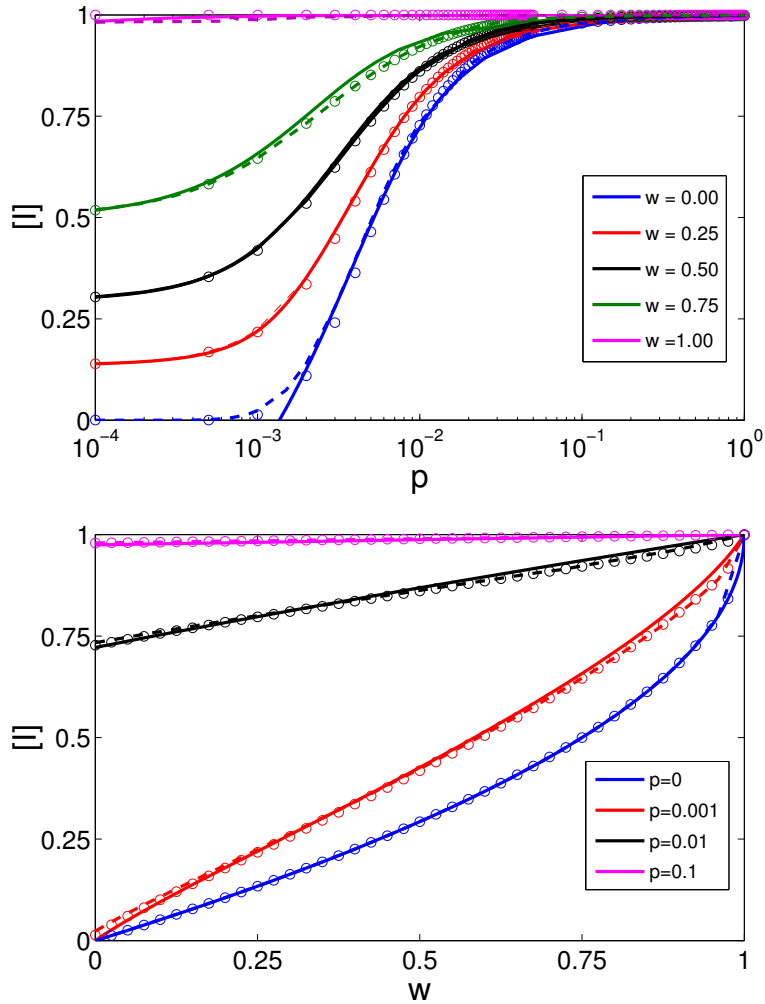}
  \caption{Dependence of the disease prevalence $[I]^{*}$ on infectiousness $p$ (top) and fraction of infected arrivals $w$ (bottom). The disease prevalence increases monotonically with infectiousness $p$ and fraction
  of infected arrivals $w$. In agent-based simulations (circles), $[I]^{*}$ is calculated over the surviving runs among $10^3$ total realizations. Homogeneous approximation (solid lines) is the analytical solution 
  of equation~(\ref{eq:S}). Heterogeneous approximation (dashed lines) is the stationary value of the numerical integration of equation~(\ref{eq:Sk}). Parameters: $r = q = 0.01$, $m_0 = 6$, $m= 5$.}
  \label{fig:epidemic_threshold}
\end{figure}

\begin{figure}[ht!]
  \bigskip 
  \centering
  \includegraphics[width=100mm,keepaspectratio]{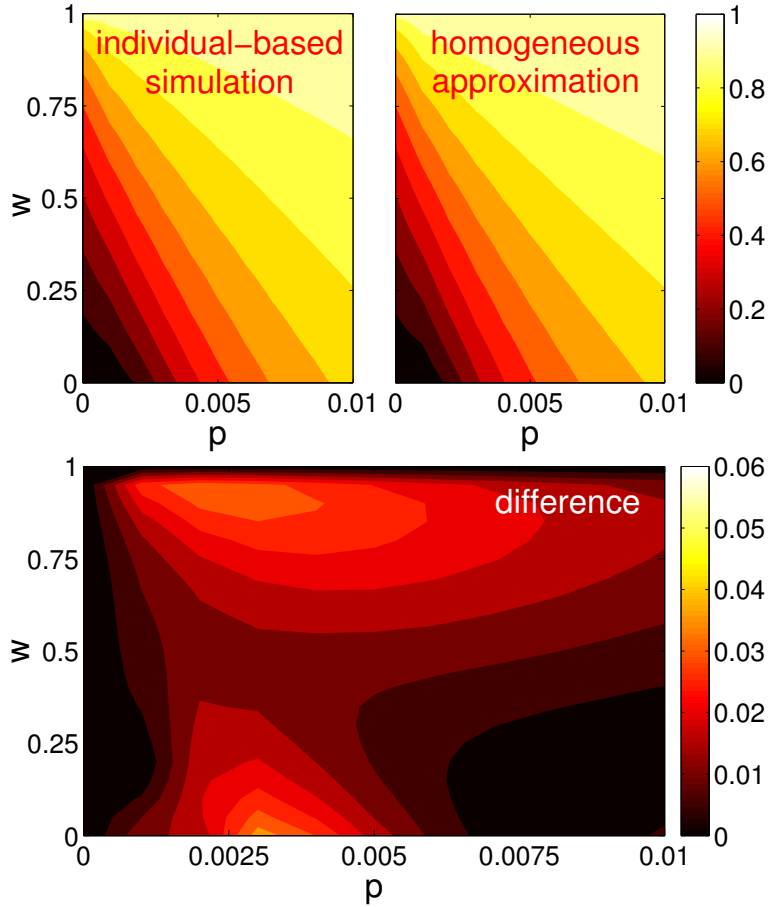}
  \caption{Performance of the homogeneous approximation in estimating the disease prevalence $[I]^{*}$ in the $(p,w)$ parameter space. Top-left: individual-based simulations. Top-right: homogeneous approximation. Bottom: absolute difference
  between individual-based simulations and the homogeneous approximation. Homogeneous approximation performs well for a large range of infectiousness $p$ and infected arrival fractions $w$. Parameters: $r = q = 0.01$, $m_0 = 6, m= 5$.}
  \label{fig:epidemic_threshold2}
\end{figure}

Let us now investigate the effect of the emergent network topology on the prevalence of the disease. 
Plots of the disease prevalence as a function of the infectiousness $p$, and the fraction of infected arrivals are shown in figures~\ref{fig:epidemic_threshold} and \ref{fig:epidemic_threshold2}. Figure~\ref{fig:epidemic_threshold} shows that the heterogeneous approximation (dashed lines) is in exact agreement with the agent-based model for the whole range of parameters $p$ and $w$. The precision of the homogeneous approximation is also high for a large range of parameter values. As illustrated in figure~\ref{fig:epidemic_threshold2}, the absolute error in estimation of the disease prevalence of the approximation is maximal for intermediate values of infectiousness $p$, but still less than $0.05$. The only qualitative discrepancy between the approximation and the agent-based model emerges at low infectiousness $p$ for zero infected arrivals ($w=0$). Here, the homogeneous model predicts the existence of an epidemic threshold, whereas in the agent-based simulation and the heterogeneous approximation the disease is found to persist for any finite infectiousness $p$.

\section{ABSENCE OF EPIDEMIC THRESHOLDS}
Summarizing the results shown so far, we can say that ongoing epidemic dynamics quickly leads to the formation of networks with finite variance. Generally, one would expect that such networks should exhibit a finite epidemic threshold.
Nevertheless, the heterogeneous approximation and simulations indicate that the epidemic can persist in these networks for any finite positive value of the infectiousness. Let us therefore investigate the apparent absence of the threshold in greater detail. 
In the following we confirm this absence by what is essentially a less naive version of the naive line of reasoning presented in the introduction.
Throughout the argument we will only consider the case where all arriving agents are susceptible, as the epidemic threshold is harder to define otherwise. 

The epidemic threshold is commonly defined as the minimal value of the infectiousness, below which a randomly picked node is susceptible with probability 1. In any finite network this implies that below the epidemic threshold each individual node is susceptible. By contrast 
in the thermodynamic limit there can still be a finite number of infected nodes as long as the density of such nodes in the network is zero, i.e. $[I]=0$. In the following we denote the state below the epidemic threshold as the disease-free state, but recognize that there may be still a finite number of infected agents.  

Because the density of infected vanishes in the disease-free state, it is apparent that the degree distribution must be independent of the 
parameters $p$ and $r$. This can also be verified in the heterogeneous approximation. Furthermore, in the thermodynamic limit, 
i.e.~$N\rightarrow \infty$, the degree distribution converges to the power law $P_{k}\sim k^{-3}$ for $k\geq m$ and thus also the degree variance of susceptibles $\sigma_{S}^{2}$ diverges in the disease-free state. Considering any state with stationary disease prevalence $[I]^{*}$, mean degree $\langle k \rangle^{*}$, susceptible mean degree and $\langle k_{S} \rangle^{*}$, the heterogeneous approximation implies
\begin{equation}
\displaystyle \frac{\rm d}{\rm dt} \langle k^{2} \rangle = q\left(m^{2}+2m-\frac{\langle k \rangle^{*}}{2}\right) + \frac{r}{2}\langle k \rangle^{*} [I]^{*} -r \left(\langle k^{2} \rangle - \langle k_{S}^{2} \rangle [S]^{*}\right).
\end{equation}
Using $\langle k^{2} \rangle = \langle k_{S}^{2} \rangle [S] + \langle k_{I}^{2} \rangle[I]$ it follows that $\sigma^{2} \rightarrow\infty$ iff $\sigma_{S}^{2} \rightarrow\infty$. In addition, the steady-state solution of the coarse-grained heterogeneous approximation provides that $1/\sigma_{S}^{2}$ vanishes only when $[I]=0$. Thus, the disease prevalence can vanish only when the degree distribution is scale-free.

The argument above provides a substantiation for our intuition that, in the disease-free state, preferential attachment will eventually lead to the formation of scale-free topologies, where the degree variance of the susceptibles diverges. Considered on its own this argument does not preclude the existence of an epidemic threshold, below which the network is disease-free and scale-free. Although we know that the epidemic threshold vanishes in models of epidemic spreading on static scale-free networks, the same is not necessarily true in the adaptive network considered here. We therefore explicitly compute the epidemic threshold by checking the stability of the disease-free state against invasion of the epidemic.   

\begin{figure}[ht!]
  \bigskip 
  \centering
  \includegraphics[width=100mm,keepaspectratio]{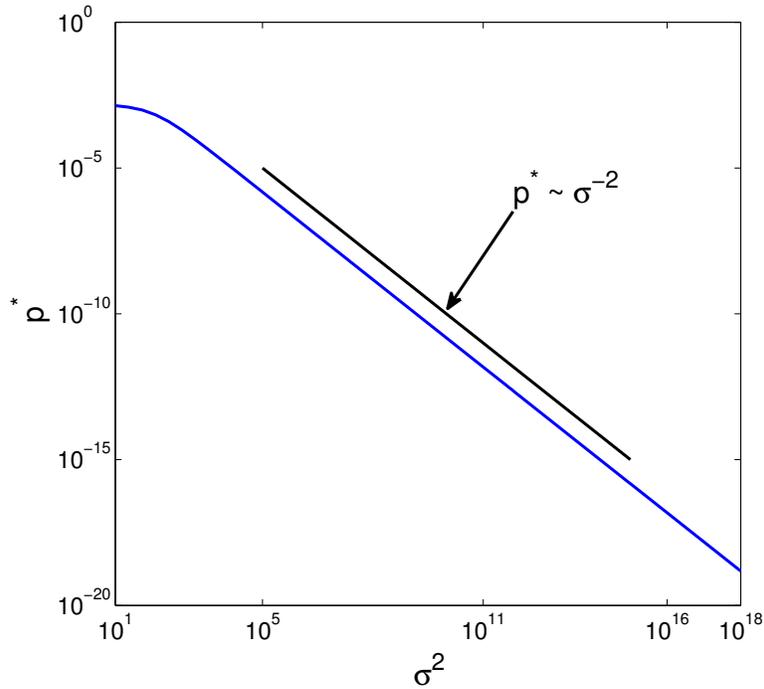}
  \caption{The epidemic threshold $p^{*}$ as a function of a given value of degree variance $\sigma^{2}$. Results have been obtained by numerical continuation of a bifurcation line from equation~(\ref{eq:low_hetero}).
 The epidemic threshold scales in the same form as on static networks, i.e. $\displaystyle p^{*} \sim {\langle k \rangle}/{\sigma^{2}}$, showing that the epidemic threshold vanishes in the thermodynamic limit. Parameters: $r = q = 0.01$, $w=0$, $m_0 = 6$, $m= 5$.}
  \label{fig:epidemic_threshold_var}
\end{figure}

In a dynamical system, a given steady state is stable if all eigenvalues of the corresponding Jacobian matrix have negative real parts.
The Jacobian matrix is obtained by a local linearisation of the system around the steady state under consideration.
For the disease-free state, we can obtain the Jacobian ${\rm \bf J}$ directly from the coarse-grained heterogeneous approximation
(equation~(\ref{eq:low_hetero})),
which yields 
\begin{eqnarray}\label{eq:jacobian}
{\rm \bf J} = \begin{pmatrix}
              \displaystyle  -q+pm-r & -p & -p+4pm \\ 
	      \displaystyle  2rm & -q-2r & 2r \\  
	      \displaystyle  \Gamma_{1} & \displaystyle \Gamma_{2}& \displaystyle \Gamma_{3}  
             \end{pmatrix}
\end{eqnarray}
where
\begin{eqnarray}
\displaystyle \Gamma_{1} & = &  \displaystyle p+\sigma_{S}^{2}+qm+2mr \nonumber \\
\displaystyle \Gamma_{2} & = &  \displaystyle -\frac{p}{2m}\sigma_{S}^{2}-r-\frac{q}{2} \nonumber \\
\displaystyle \Gamma_{3} & = &  \displaystyle \frac{p}{2m}\sigma_{S}^{2}+r-\frac{q}{2}.
\end{eqnarray}
We can now compute the epidemic threshold $p^{*}$ as the critical threshold in $p$ where the disease-free state loses its stability through a transcritical bifurcation. Since the Jacobian ${\rm \bf J}$ depends only on the value of the susceptible degree variance $\sigma_{S}^{2}$
but not on its partial derivatives, we can treat the susceptible degree variance $\sigma_{S}^{2}$ as a parameter and obtain the epidemic threshold as its function by numerical continuation of the bifurcation point. For 
sufficiently large degree variance $\sigma_{S}^{2}$, we find the scaling relation $\displaystyle p^{*}\sim {\langle k \rangle}/{\sigma_{S}^{2}}$ (figure~\ref{fig:epidemic_threshold_var}), which recovers a famous result from epidemics on static networks. The epidemic threshold thus vanishes in the thermodynamic limit when $\sigma_{S}^{2} \rightarrow \infty$. 

The two paragraphs above provide a clearer picture of the paradoxical situation sketched already in the introduction. 
If the disease prevalence vanishes then the network becomes scale-free, however the scale-free network has a vanishing epidemic 
threshold and thus supports finite prevalence. In principle this paradox can be resolved in two ways, a) either the network 
does not become stationary but switches back and forth between scale-free and exponential phases, or b) the network approaches a state 
where the epidemic persists in an exponential topology. 

It is apparent that in finite networks the temporal dynamics can lead to a disease-free scale-free state. In this case the network goes through an initial exponential phase where an epidemic threshold exists that leads to the extinction of the epidemic. Subsequently, 
a scale-free topology is build up, in which the epidemic threshold vanishes. However the epidemic cannot reappear as no infected are left 
which could reignite the epidemic. 

\begin{figure}[ht!]
  \bigskip 
  \centering
  \includegraphics[width=100mm,keepaspectratio]{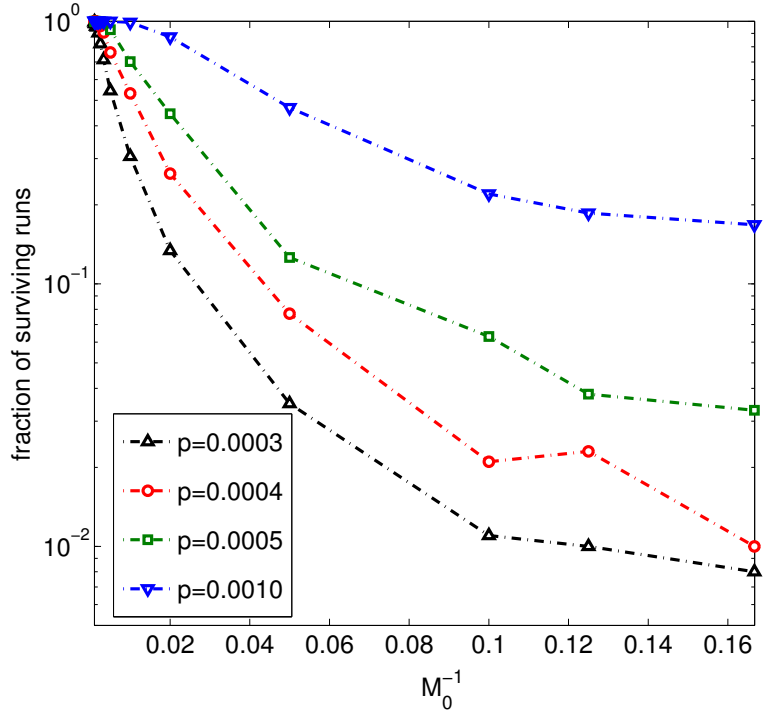}
  \caption{The fraction of surviving runs for low infectiousness $p$ as a function of the initial network size $M_{0}$. The fraction of surviving runs increases as the 
  initial network size $m_{0}$ increases. In order to ensure the same initial average degree $m$, the Barab\'{a}si-Albert growth model with $m=5$ and $m_{0}=6$ is iterated until the 
  network reaches size $M_{0}$. Then nodes are assigned states and the infection, removal, and network growth processes take place simultaneously. Parameters: $r = q = 0.01$, $w=0$, $m_0 = 6$, $m= 5$.}
  \label{fig:survivefrac}
\end{figure}

We emphasize that the mechanism above is a finite-size effect. In the thermodynamic limit there would generally be some infected left even in the disease-free state. To see this consider that infected are removed exponentially, 
so that after any finite time there will still be infected left which can launch the network back into an endemic state, such that scenarios a) and b) are the only possible outcomes.
The finite-size extinction of the epidemic is demonstrated explicitly in figure~\ref{fig:survivefrac}, which shows that 
the fraction of agent-based simulation runs in which the epidemic persists increases with the initial network size $M_{0}$.  

\begin{figure}[ht!]
  \bigskip 
  \centering
  \includegraphics[width=100mm,keepaspectratio]{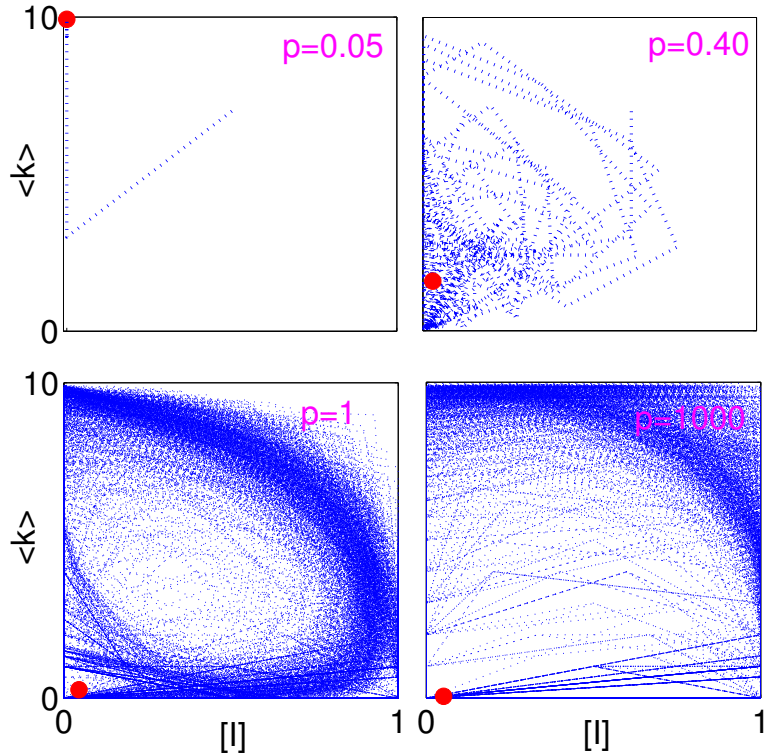}
  \caption{The dynamical evolution in the $[I]-\langle k \rangle$ plane for high removal rate $r$. At high infectiousness $p$, the observed behaviour resembles a homoclinic trajectory. The network grows from a 
  healthy initial state until an infected eventually arrives and the disease quickly spreads over the network and the infected individuals subsequently die. Homogeneous approximation (red circles) fails to capture this behaviour and 
  predicts a stable equilibrium. Parameters: $q = 0.01$, $r = 0.10$, $m_0 = 6$, $m= 5$, $w=0.1$.}
  \label{fig:approximation_failure}
\end{figure}

Precluding the finite-size effect described above, simulation runs in the parameter range considered so far approach a finite prevalence for any positive value of infectiousness. The paradox outlined above is thus resolved by alternative a) the model does not have an epidemic threshold although the evolving topologies remain exponential. As a final step in our exploration we now search for evidence of the temporal resolution of the paradox, the alternative b).

Up to now we have considered relatively small removal rates $r$. At sufficiently high removal rate, the evolution of the disease and the topology exhibits dynamics different from the convergence to a stationary 
state. Figure~\ref{fig:approximation_failure} shows a representative evolution at non-zero fraction of infected arrivals $w$. At the combination of high infectiousness $p$ and high removal rate $r$, the dynamics resembles 
a homoclinic trajectory in the $\langle k \rangle - [I]$ plane. The disease spreads quickly over the network and it covers the whole population immediately. Then disease-induced removals dominate and the population becomes extinct until 
new healthy individuals arrive and the disease spreading restarts, which leads to cycles of population growth and collapse. 

When all arrivals are susceptible, i.e. $w=0$, the epidemics in the finite population disappears entirely and the collapse-and-growth cycle cannot be completed. A disease-free scale-free network then emerges. 
Based on the arguments above one can suspect that the same behaviour cannot occur in the thermodynamic limit. 
Instead it is likely that the observed dynamics forms part of a homoclinic cycle, where long phases of very low disease prevalence are disrupted by sharp outbreaks. Similar dynamics have for instance been studied in an adaptive-network model of cooperation among agents \cite{Zschaler2010}. In the present model, the moment closure approximation fails to capture this behaviour and predicts a stable steady-state solution for high values of $r$. 

The observed dynamics at high removal rates would correspond to diseases with very high mortality where infected individuals die almost immediately. This is reminiscent of the massive pandemics in history, where humanity was exposed to new pathogens with very high virulence. While such diseases could lead to the deaths of large fractions of populations, continuous supply of healthy individuals through immigration provided new hosts to the disease and 
introduced new bursts of disease spreading which caused repeated epidemics cycles.   

\section{DISCUSSION}\label{sec:discussion}
In the present paper we have investigated the dynamics of a fatal disease in a growing population. 
Our main finding is that no epidemic threshold exists in this model although the variance of the degree distribution remains finite.
Thus even in evolving exponential network topologies ``unlikely" diseases with very low infectiousness can persist indefinitely. 

We presented a detailed analytical exploration for the case of low removal rate, where the prevalence of the disease reaches a stationary level. In the growing population the ongoing epidemic dynamics eliminates the nodes of high degree and thus leads to the formation of 
exponential topologies for which the variance of the degree distribution is finite. However, this mechanism only reduces the width of the 
degree distribution so far that the epidemic can still persist. Thus for any finite value of the infectiousness the network adapts its topology such that the variance of the degree distribution is lower to point where the epidemic can still be sustained, which explains the observed absence of the epidemic threshold.  

When the removal rate is sufficiently high, we observed a much more dynamic picture, where degree variance acts as a supply of fuel that slowly builds up before being burned in large epidemic outbreaks. These dynamics are reminiscent of the massive deaths from epidemic diseases in newly forming urban areas. A detailed analysis thus appears as a promising target for future work.
  
The dynamical feedback between the population structure and the epidemic disease has been so far studied in a number of articles in the past five years \cite{Gross2006,Shaw2008,Risau-Gusman2009,Graeser2011,Wang2011,Marceau2010}. However, models captured mostly social interactions in non-fatal diseases. 
A significant obstacle to progress in this line of work is that the network evolution in most models is driven by behavioural changes of individuals (whom to meet, how often to wash hands). However, for behaviour often no records exist such that model predictions cannot easily be compared with real world data. 
By contrast, in the model proposed here, the social network evolves due to demographic processes such as migration and death on which data may be easier to obtain.  
   
We believe that the proposed model is relevant for epidemics in rapidly growing cities, especially in the developing countries.
We hope that in this context connecting the model to real world data will be feasible in the future. One can easily 
imagine extensions of the present model that incorporate policy measures such as vaccination, quarantine, or regulation of migration. 
We hope that this will in the future lead to the formulation of more efficient policies for combating epidemic diseases. 

\bibliographystyle{}

\end{document}